\documentstyle[aps,preprint,tighten,epsfig]{revtex}

\newbox\rotbox
\newcommand{\be}{\begin{eqnarray}}
\newcommand{\ee}{\end{eqnarray}}
\newcommand\tag{\hbox to hsize}

\def\mytoday#1{{}\ifcase\month\or
January\or February\or March\or April\or May\or June\or
July\or August\or September\or October\or November\or December\fi
\space \number\year}

\begin{document}
\draft
\title{$K^{\ast }$ nucleon hyperon form factors and nucleon strangeness}
\author{Hilmar Forkel}
\address{Institut f{\"u}r Theoretische Physik, Universit\"{a}t Heidelberg, \\
Philosophenweg 19, D--69120 Heidelberg, Germany}
\author{Fernando S. Navarra and Marina Nielsen}
\address{Instituto de F\'{\i}sica, Universidade de S\~ao Paulo, \\
C.P. 66318, \\
05315-970 S\~ao Paulo, SP, Brazil}
\maketitle
\date{\today}

\begin{abstract}
A crucial input for recent meson hyperon cloud model estimates of the
nucleon matrix element of the strangeness current are the nucleon--hyperon--$%
K^{\ast }$ ($NYK^{\ast }$) form factors which regularize some of the arising
loops. Prompted by new and forthcoming information on these form factors
from hyperon--nucleon potential models, we analyze the dependence of the
loop model results for the strange--quark observables on the $NYK^{\ast }$
form factors and couplings. We find, in particular, that the now generally
favored soft $N\Lambda K^{\ast }$ form factors can reduce the magnitude of
the $K^{\ast }$ contributions in such models by more than an order of
magnitude, compared to previous results with hard form factors. We also
discuss some general implications of our results for hadronic loop models.
\end{abstract}



\preprint{HD-TVP-, U of USP-XXX}


\narrowtext

\section{Introduction}

Understanding the non--valence quark content of hadrons remains, despite a
history spanning over two decades, a major theoretical challenge. The
interest in non--valence physics derives mainly from the unique
opportunities which it provides for new insights into quantum aspects of
hadron structure beyond the naive and spectroscopically very successful
quark model. Important questions in this realm include\ modifications of the
QCD vacuum inside hadrons \cite{don86}, the mechanism of flavor mixing and
the origin of the OZI--rule \cite{gei91,gei97}, the structure of constituent
quarks \cite{kap88}, and the role of gluons in the dynamics of the
(isoscalar) strange--quark sea in the nucleon \cite{dos98}.

The strangeness distribution inside of the nucleon \cite{dec93} represents
the most intensely studied example of a hadronic non--valence quark effect.
Currently the vector channel of this distribution, as described by the
strange vector form factors, is a focus of experimental \cite{mit,Mai,TJN}
and theoretical \cite{dec93,mus94} research. Since systematic and
model--independent approaches have still little predictive power for the
nucleon's strangeness content (as exemplified by studies in chiral
perturbation theory \cite{bor97,ram97,hem98,oh99} and on the lattice \cite
{lat}), most previous and current theoretical analyses of the strangeness
form factors were model--based.

Among the first and most transparent models for the vector form factors were
those which implement a kaon--cloud of the nucleon \cite{had} and thus
complement pole dominance approaches \cite{pole}. In kaon--cloud models the
nucleon's strangeness distribution is generated by fluctuations of the
``bare'' (i.e. nonstrange) nucleon into kaon--hyperon intermediate states
which are described by the corresponding one--loop Feynman graphs \cite{had}%
. The two crucial assumptions underlying the loop model are 1) that the
lightest valence--strangeness carrying intermediate states generate the
dominant contribution to the strangeness content and hence give at least a
rough estimate of its size, and 2) that rescattering (i.e. multi--loop)
contributions are suppressed (despite large couplings).

Both of these assumptions have recently been challenged. A dispersive
analysis on the basis of analytically continued $K-N$ scattering data
demonstrated that rescattering corrections are important even at low
momentum transfers, both to restore unitarity and to build up resonance
strength in the $\phi $ meson region \cite{ram97c}. Furthermore, a study in
an ``unquenched'' quark model found the contributions from higher--lying
intermediate states (up to surprisingly large invariant masses)
indispensable for the calculation of the strange quark distribution \cite
{gei97} and prompted our collaborators and us to investigate these issues in
a complementary hadronic one--loop model\ \cite{bar98}. In addition to the
original $K-Y$ loops ($Y=\Lambda ,\Sigma $), we included the next
higher--lying intermediate states, i.e. the $K^{\ast }-Y$ pairs. In the
first part of this study the corresponding loops were evaluated using
Bonn--J\"{u}lich $K^{\ast }NY$ form factors (see below),\ as in the original 
$K-Y$ model. The results were discouraging: the $K^{\ast }$ contributions
were found to be larger than those from the kaon loop, and their dispersive
analysis indicated strong unitarity violations.

The anomalously large and apparently unrealistic $K^{\ast }$ contributions
could be traced to the large $K^{\ast }$ tensor couplings and, in
particular, to the very large cutoff parameter of the $K^{\ast }N\Lambda $
form factor taken from the Bonn--J\"{u}lich potential model \cite{hol89}.
Since both the Nijmegen potential \cite{nag77,rij98} and the forthcoming
update of the Bonn--J\"{u}lich $NY$ potential\ \cite{hai98} find
substantially smaller values for this cutoff, we feel that a detailed and
quantitative analysis of \ the cutoff (and coupling) dependence of the $%
K^{\ast }$ contributions (covering the whole range of so far proposed
values) \ would be a useful contribution to the ongoing discussion \cite
{mel97,mel99} of hadron loop model applications to nucleon strangeness. Such
an analysis is one of the main objectives of the present paper. Furthermore,
we will calculate and discuss the momentum dependence of the strange vector
form factors in the loop model at low momentum transfers ($Q^{2}\leq 1$ GeV)
relevant for the present and planned measurements at MIT-Bates, TJNAF and
MAMI.

\section{The meson-hyperon loop model}

\label{mod}

We begin by recapitulating the definition of the strange vector form
factors, the pertinent features of the hadronic loop model \cite{had}, and
its extension to additional intermediate states containing $K^{\ast }$
mesons \cite{bar98}. The focus of our investigations will be on the nucleon
matrix element of the strangeness current, which is parametrized by two
invariant amplitudes, the Dirac and Pauli strangeness form factors $%
F_{1,2}^{(s)}$, 
\begin{equation}
\langle N(p^{\prime })|\bar{s}\gamma _{\mu }s|N(p)\rangle ={\bar{U}}%
(p^{\prime })\left[ F_{1}^{(s)}(q^{2})\gamma _{\mu }+i{\frac{\sigma _{\mu
\nu }q^{\nu }}{2m_{N}}}F_{2}^{(s)}(q^{2})\right] U(p)\ .
\end{equation}
Here $U(p)$ denotes the nucleon spinor and $F_{1}^{(s)}(0)=0$, due to the
absence of an overall strangeness charge of the nucleon. The leading
nonvanishing moments of these form factors are the Dirac and Sachs (square)\
strangeness radii 
\begin{equation}
\langle r_{s}^{2}\rangle _{D}=\left. 6{\frac{d}{dq^{2}}}F_{1}^{(s)}(q^{2})%
\right| _{q^{2}=0},\qquad \quad \langle r_{s}^{2}\rangle _{S}=\left. 6{\frac{%
d}{dq^{2}}}G_{E}^{(s)}(q^{2})\right| _{q^{2}=0},
\end{equation}
as well as the strangeness magnetic moment 
\begin{equation}
\mu _{s}=F_{2}^{(s)}(0)\,.
\end{equation}
(The Sachs radius is obtained from the electric Sachs form factor $%
G_{E}^{(s)}(q^{2})=F_{1}^{(s)}(q^{2})+q^{2}/(4m_{N}^{2})\,F_{2}^{(s)}(q^{2})$
and related to the Dirac radius by $\langle r_{s}^{2}\rangle _{S}=\langle
r_{s}^{2}\rangle _{D}+3\mu _{s}/(2m_{N}^{2})$.)

As mentioned above, we have chosen a hadronic one--loop model containing $K$
and $K^{\ast }$ mesons as the dynamical framework for the calculation of
these moments. This model is based on the meson--baryon effective
lagrangians 
\begin{eqnarray}
{\cal L}_{MB} &=&-g_{ps}\bar{B}i\gamma _{5}BK\ \ \ ,  \label{1aa} \\
{\cal L}_{VB} &=&-g_{v}\left[ \bar{B}\gamma _{\alpha }BV^{\alpha }-\frac{%
\kappa }{2m_{N}}\bar{B}\sigma _{\alpha \beta }B\partial ^{\alpha }V^{\beta }%
\right] \;,  \label{la}
\end{eqnarray}
where $B$ ($=N,\Lambda ,\Sigma $), $K$, and $V^{\alpha }$ are the baryon,
kaon, and $K^{\ast }$ vector-meson fields, respectively, $m_{N}=939$ MeV is
the nucleon mass and $\kappa $ is the ratio of tensor to vector coupling, $%
\kappa =g_{t}/g_{v}$. In order to account for the finite extent of the above
vertices, the model includes form factors from the Bonn--J\"{u}lich $N-Y$
potential \cite{hol89} at the hadronic $KNY$ and $K^{\ast }NY$ ($Y=\Lambda
,\Sigma $) vertices, which have the monopole form 
\begin{equation}
F(k^{2})=\frac{m^{2}-\Lambda ^{2}}{k^{2}-\Lambda ^{2}}  \label{ff}
\end{equation}
with meson momenta $k$ and the physical meson masses $m_{K}=495$ MeV and $%
m_{K^{\ast }}=895$ MeV \cite{PDG}. These form factors render all encountered
loop integrals finite and reproduce the on--shell values of the mesonic
couplings. The range of currently favored values for the couplings $%
g_{ps},g_{v},\kappa $ and cutoff parameters $\Lambda _{K}$ and $\Lambda
_{K^{\ast }}$, as well as their impact on the strangeness observables, will
be discussed below.

Since the non-locality of the meson-baryon form factors (\ref{ff}) gives
rise to vertex currents, gauge invariance was maintained in \cite{bar98} by
introducing the photon field via minimal substitution in the momentum
variable $k$ \cite{ohta}. (Consequences of the non--uniqueness of this
prescription are discussed in Refs. \cite{ohta,bos92,wan96}.) The resulting
nonlocal seagull vertices are given explicitly in \cite{bar98}.

The diagonal couplings of $\bar{s}\gamma _{\mu }s$ to the strange mesons and
baryons in the intermediate states are straightforwardly determined by
current conservation, i.e. they are given by the net strangeness charge of
the corresponding hadron. The situation is more complex for the
non--diagonal (i.e. spin--flipping) coupling $F_{KK^{\ast }}^{(s)}(0)$ of
the strange current to $K$ and $K^{\ast }$, which is defined by the
transition matrix element

\begin{equation}
\langle K_{a}^{\ast }(k_{1},\varepsilon )|{\overline{s}}\gamma _{\mu
}s|K_{b}(k_{2})\rangle =\frac{F_{KK^{\ast }}^{(s)}(q^{2})}{m_{K^{\ast }}}%
\,\epsilon _{\mu \nu \alpha \beta }\,k_{1}^{\nu }\,k_{2}^{\alpha
}\,\varepsilon ^{\ast \beta }\,\delta _{ab}\;  \label{spinfl}
\end{equation}
(where $a$ and $b$ are isospin indices and $\varepsilon ^{\beta }$ is the
polarization vector of the $K^{\ast }$). This coupling was estimated in \cite
{bar98} on the basis of the vector meson dominance model of Ref. \cite{gm},
with the result $F_{KK^{\ast }}^{(s)}(0)=1.84$.

The relevant one--loop Feynman graphs of the model are completely determined
by the above vertices and the standard meson and baryon propagators. The
diagrams containing strange mesons fall into three categories, corresponding
to fluctuations of the nucleon into either $K-Y$ or $K^{\ast }-Y$ pairs, or
to the strangeness--current induced spin flip transition from a $K-Y$ to a $%
K^{\ast }-Y$ intermediate state. The explicit expressions for the
corresponding loop amplitudes are collected in Appendix A. In the following
calculations we will concentrate on the $\Lambda $ contributions and omit
the comparatively negligible $\Sigma $ contribution \cite{had,bar98}.

\section{Internal hadronic vertices}

In this section we investigate the dependence of the loop-model results on
the couplings and cutoffs which parametrize the internal $K(K^{\ast
})\Lambda N$ vertices. In our previous analysis \cite{bar98} we have used
the ($SU(3)$ based) couplings $g_{ps}/\sqrt{4\pi }=-3.944$, $g_{v}/\sqrt{%
4\pi }=-1.588$, $\kappa =3.26$ and the cutoff parameter values $\Lambda
_{K^{\ast }}=2.2$ (2.1), $\Lambda _{K}=1.2(1.4)$ GeV of the Bonn--J\"{u}lich 
$NY$ potential \cite{hol89}. The cutoffs were determined from
hyperon--nucleon scattering data, with the numbers in parenthesis denoting
values obtained in an alternative model for the baryon-baryon interaction.

The numerical results of the loop model \cite{bar98} are summarized in Table
I. A glance at these numbers shows that the magnitude of the $K^{\ast }$
contributions exceeds those of the $K$ contributions by factors of $5-10$.
As already mentioned, the main reason for these unrealistically large
contributions can been traced to the unusually large $K^{\ast }N\Lambda $
cutoff parameter $\Lambda _{K^{\ast }}=2.2$ GeV found in \cite{hol89}. It
has twice the size of the typical hadronic scale $\sim 1$ GeV around which
such cutoff parameters lie normally. A substantially larger value (which in
our case also exceeds the largest hadron masses in the loops by a factor of
two) must be considered suspect in any model with hadronic degrees of
freedom since one expects their quark--gluon substructure to become relevant
at such scales.

Indeed, the appearance of anomalously large cutoffs in a potential model
suggests that those cutoffs are burdened with\ short--distance physics (not
directly related to the $K^{\ast }$ sector)\ which would otherwise remain
unaccounted for. Literally taking such effects over to the loop--model
estimates of the nucleon's strangeness content, by fully associating them
with the physical $K^{\ast },$ would therefore very likely be misleading. A
hint that the $K^{\ast }$ sector of the original Bonn--J\"{u}lich potential
might indeed be overburdened can be obtained from a comparison with the
conceptually similar Nijmegen $NY$ potential \cite{nag77,rij98}. The
Nijmegen potential contains more degrees of freedom in the scalar meson
sector (including a pomeron) and finds a much smaller $K^{\ast }$ cutoff $%
\Lambda _{K^{\ast }}\simeq 1.2$ GeV. Morevoer, a smaller $K^{\ast }$ cutoff
is also favored in the forthcoming update of the Bonn--J\"{u}lich potential 
\cite{hai98}. The $K^{\ast }$ cutoff of this model is expected to lie around 
$1.5$ GeV \cite{mel299}.

Hence, nowadays smaller cutoffs seem to be consistently favored by $NY$
potential models. This motivated our reanalysis of the loop--model results
of Ref. \cite{bar98} for the nucleon's strangeness observables (as also
suggested in \cite{mel99}) which we will discuss below. To get a qualitative
idea of the range of coupling and cutoff values to be considered, and to
maintain the underlying philosophy of the loop model, we will orient
ourselves at the values used in the existing potential models. Of course,
these values depend to some extent on the particular dynamics and particle
content of a given model, and since the various potential models and also
our model differ in this respect, the corresponding  parameter sets cannot
be directly compared or strictly related to each other. However, we expect
the parameter ranges considered below, in particular for the cutoffs, to
cover most of the physically reasonable parameter space of our model.

We begin our discussion of the parameter dependence by documenting the
sensitivity of the results to variations in the cutoff. To this end, we
display the cutoff dependence in the range $1{\rm GeV}\leq \Lambda _{K^{\ast
}}\leq 2.5\,{\rm GeV}$, which covers all so far proposed values for $\Lambda
_{K^{\ast }}$. The first two figures show the behavior of the Dirac
strangeness radius (Fig. 1a) and the strangeness magnetic moment (Fig. 1b)
as a function of $\Lambda _{K^{\ast }}$, with the $K^{\ast }$ couplings
fixed at the old, $SU(3)$--based Bonn--J\"{u}lich values given above. The
full curves show the total results, the dashed ones represent the $K^{\ast
}K^{\ast }\Lambda $ contributions, the dash--dotted ones the $KK^{\ast
}\Lambda $ contributions, and the dotted curve corresponds to the result
from the $K$ loop. Clearly, the cutoff dependence is very pronounced,
reducing e.g. the value of Ref. \cite{bar98} for the magnetic moment by
almost an order of magnitude for $\Lambda _{K^{\ast }}\simeq 1.5$ GeV. The
reduction is even considerably stronger for the Nijmegen value $\Lambda
_{K^{\ast }}\simeq 1.2$ GeV.

Although a strong cutoff dependence was to be expected (especially due to
the enhanced degree of divergence of $\ $the $K^{\ast }$ loops with
derivative couplings), its actual magnitude is still surprising. If cutoff
sizes of the order of that used in the Nijmegen potential were to be the
most realistic, some of the conclusions reached in \cite{bar98} would have
to be revised. In particular, for cutoffs of the Nijmegen size some sort of
``convergence'' of the intermediate--state sum (to one loop)\ could not
anymore be excluded. In this case the kaon cloud contribution might well be
sufficient for a first orientation about the overall size of the nucleon's
strangeness content in hadronic one--loop models, as advocated in \cite
{mel99} and in contrast to the findings of Ref. \cite{gei97} in the quark
model. Other problematic aspects of hadron--loop models, such as the
expected importance of rescattering corrections and unitarity violations 
\cite{ram97c}, are of course not affected by these arguments.

The strong cutoff dependence of the strangeness observables exposes another,
both conceptual and practical problem of hadronic loop models. Although the
cutoff is principally a physical parameter (which indicates up to which
resolution a purely hadronic description might be adequate), the foundations
of the approach are not solid enough to give it a precise and quantitative
meaning. Moreover, the available $NY$ scattering data do not allow an
accurate determination of the $K^{\ast }NY$ form factors even in the
framework of a specific potential model, not to mention other sources of
uncertainty as, for example, the largely uncontrolled off--shell ambiguities
incurred by transplanting such form factors into another model context. As a
consequence, the numerical value of the $K^{\ast }$ cutoff cannot be
accurately determined, and the corresponding uncertainty propagates,
amplified by a hightened sensitivity, into the strangeness observables.
These facts already imply that at most semi--quantitative predictions can be
expected from the hadron--loop approach.

The existing $NY$ potential models differ not only in the momentum
dependence of the $K^{\ast }NY$ form factors, but also in the values of the
corresponding $K^{\ast }$ couplings. The $K^{\ast }$ vector (Dirac) coupling
of the Nijmegen potential is, for example, considerably smaller than the one
of Ref. \cite{hol89} which we used above. In order to illustrate the
dependence of the strange form--factor moments on these couplings, we
compare in Fig. 2 the $K^{\ast }$ contributions to the strangeness radius
(Fig. 2a) and magnetic moment (Fig. 2b)\ for the five pairs ($-1.588,3.26$),
($-0.8,2.0$), ($-0.8,4.0$), ($-2.0,2.0$), ($-2.0,4.0$) of coupling values ($%
g_{v}/\sqrt{4\pi },\kappa $). The first of these pairs corresponds to the
values of Holzenkamp et al. while the others were chosen to encompass the
range of values which appear in other potential models. The Figs. 2 a,b
demonstrate that the variations of the strangeness observables due to
different choices for the couplings can be quite substantial for large
values of the $K^{\ast }$ cutoff. The sensitivity to the couplings remains
fairly small, however, for $\Lambda _{K^{\ast }}$ of the order of $1$ GeV,
in particular for the strange magnetic moment and for couplings in the range
between the Bonn-J\"{u}lich ($-1.588,3.26$) and Nijmegen \cite{rij98} ($%
-1.45,2.43$) values\footnote{%
These Nijmegen couplings are obtained from Table II of Ref. \cite{rij98} by
using the SU(3) relations given in their Eq. (2.14).}.

Throughout all of the above calculations we have kept the values of Ref. 
\cite{hol89} for the cutoff and coupling of the kaon fixed. The Nijmegen
potential uses a pseudovector coupling which cannot be related without
off--shell ambiguities to the pseudoscalar coupling of the Bonn--J\"{u}lich
potential. On--shell, the equivalent pseudoscalar coupling of the Nijmegen
potential ($g_{ps}/\sqrt{4\pi }\simeq -4.0$) differs by less than 5\% from
the Bonn--J\"{u}lich coupling used here, and the Nijmegen cutoff parameter $%
\Lambda _{K}\simeq 1.28$ GeV \cite{rij98} is similarly close to the one we
use. In any case, the cutoff and coupling values from potential models
should not be taken too literally since, as discussed above, the limited
available data and the implicit model assumptions do not allow their precise
(and unique) determination.

We close this section with a remark on alternative choices for the internal
form factors. The Bonn monopole form factors (\ref{ff}) have two well-known
limitations: an artificial zero for cutoffs of the size of the meson mass
(due to on-shell normalization) on which we will comment in the next
section, and an unphysical singularity at time-like momenta $q^{2}=\Lambda
^{2}$. Although the singularity can be tamed by the usual causal boundary
condition, it will still affect the values of the loop amplitudes to a
certain extent. This problem could be avoided, although at the price of
additional model dependence, by choosing alternative, singularity-free
extrapolations of the Bonn form factors into the time-like region. One such
extrapolation was proposed in Ref. \cite{fri99}.

\section{Momentum dependence of the strange form factors}

All of the currently running or planned experiments measure the strange
vector form factors at $Q^{2}\neq 0$. Therefore, their data need to be
extrapolated in order to obtain the leading form-factor moments. This
extrapolation requires information on the momentum dependence of the form
factors (at low energies) and thus on the spatial strangeness distribution
inside the nucleon. In the following section we will evaluate the loop-model
predictions for the momentum dependence by extending the above calculations
to momentum transfers $Q^{2}\leq $ 1 GeV$^{2}.$ Both strange vector form
factors will be measured in about the same $Q^{2}$-range by the planned
spectrometer experiment G0 at Jefferson Lab \cite{TJN}. \ 

In the loop model, the coupling of the strange current to hyperon, meson, $%
K(K^{\ast })NY$ vertex or $K/K^{\ast }$ transition vertex inside the loop is
described by corresponding vertex functions, which we list in the appendix.
These vertex functions contain the full momentum dependence of the form
factors and can be evaluated numerically. Fixing the cutoffs for the $%
K^{\ast }$ and the kaon at $\Lambda =1.2$ GeV and using the Nijmegen values
for the coupling, we show in Fig. 3 (solid line) the resulting strange
magnetic form factor 
\begin{equation}
G_{SAMPLE}^{(s)}(Q^{2})=G_{M}^{(s)}(Q^{2})=F_{1}^{(s)}(Q^{2})+F_{2}^{(s)}(Q^{2}).
\end{equation}
together with the data point measured by the SAMPLE experiment at $Q^{2}=0.1$
GeV$^{2}$ \cite{mit}. (For comparison we have also included the older data
point from the first runs only.)\ Figure 4 (solid line) shows the loop-model
prediction for the combination 
\begin{equation}
G_{HAPPEX}^{(s)}(Q^{2})=G_{E}^{(s)}(Q^{2})+0.39G_{M}^{(s)}(Q^{2}),
\end{equation}
measured by the HAPPEX\ collaboration, together with their data point at $%
Q^{2}=0.48$ GeV$^{2}$ \cite{TJN}. The corresponding loop-model results for
the Dirac strangeness radius and magnetic moment are given in Table I.

At first sight, the results are in disagreement with both the SAMPLE\ and
the HAPPEX data. The measured form factor combinations are both positive (at
their respective momentum transfers) while the loop model leads to negative
form factors of smaller slopes and magnitudes. In particular, the loop model
predicts - like the majority of the other so far investigated models - a
small and negative value for the strangeness magnetic moment.

With regard to the SAMPLE result, however, there is a caveat. The derivation
of the strangeness form factor from the measured asymmetry requires
knowledge of the neutral weak axial form factor $G_{A}^{\left( Z\right) }$
of the nucleon and, in particular, its isovector radiative correction \cite
{mus94}. The newest published value $G_{SAMPLE}^{(s)}(Q^{2}=0.1$ GeV$%
^{2})=0.61\pm 0.17\pm 0.21$ is based on the theoretical estimate \cite{mus90}
for this radiative correction, which carries a substantial amount of
uncertainty. The SAMPLE collaboration is currently measuring with a
deuterium target \cite{bat94} in order to pin down the axial weak form
factor contribution separately. Only after taking these data or more
accurate and reliable theoretical estimates into account (which could
substantially change the quoted value for $G_{SAMPLE}^{(s)}(Q^{2}=0.1$ GeV$%
^{2})$ both in sign an magnitude) can the strange magnetic moment be
extracted unambigously and compared to the small value of the loop-model
prediction.

Furthermore, it is interesting to note that the results of the loop model
could be made consistent with the HAPPEX data by allowing for smaller cutoff
values, as suggested by recent applications of the meson cloud model to
deeply inelastic scattering \cite{mst,cdnn}. In order to illustrate this
point, we show in Figs. 3 and 4 (dashed lines) the results obtained by using 
$\Lambda =0.9$ GeV, which also brings the magnetic form factor somewhat
closer to the SAMPLE data point. Moreover, this cutoff value is very close
to the $K^{\ast }$ mass and therefore effectively switches off the internal $%
K^{\ast }$ form factors (cf. Eq. (\ref{ff})). As a consequence, the
contributions from the $K^{\ast }$ and the $K/K^{\ast }$ transition are
completely negligible relative to the kaon contribution. This shows that the 
$K^{\ast }$ contributions worsen the agreement of the loop model predictions
with the current experimental results. However, the zeros of the Bonn form
factors have to be regarded as artefacts of the on-shell normalization and
the results for the small cutoffs should therefore be viewed with caution.
The latter reservations are enhanced by the difficulties with the physical
interpretation of cutoffs of the same size as particle masses in a loop.

In Figs. 5 and 6 we show the $K$, $K^{\ast }$ and $K/K^{\ast }$ transition
contributions to the Dirac ($F_{1}^{(s)}$) and Pauli ($F_{2}^{(s)}$) form
factors separately (again for $\Lambda =1.2$ GeV and Nijmegen couplings).
These figures demonstrate that even with the new values for cutoffs and
couplings adopted in the present paper, the $K^{\ast }$ and $K/K^{\ast }$
contributions are still of the same order of magnitude as the kaon
contributions. This is in contrast to the much smaller size of these
contributions to the strangeness magnetic moment in the chiral quark model 
\cite{han99}. In the latter the tensor coupling of the $K^{\ast }$ to the
quarks is small, while the contributions from the vector coupling partially
cancel each other and become proportional to the difference between the
light and strange quark constituent masses, and therefore also small.

Finally, we present the loop-model result for the form-factor combination 
\begin{equation}
G_{MAMI}^{(s)}(Q^{2})=G_{E}^{(s)}(Q^{2})+0.22G_{M}^{(s)}(Q^{2})
\end{equation}
which the forthcoming MAMI A4 experiment \cite{Mai} plans to measure at $%
Q^{2}=0.23$ GeV$^{2}$. We find 
\begin{equation}
G_{{\scriptsize \mbox{MAMI}}}^{(s)}(Q^{2}=0.23\mbox{GeV}^{2})=\left\{ 
\begin{tabular}{c}
$-0.012\;\mbox{
for }\Lambda =0.9\mbox{ GeV}$ \\ 
$-0.046\;\mbox{
for }\Lambda =1.2\mbox{ GeV}$%
\end{tabular}
\right. 
\end{equation}
where the same reservations as above apply to the smaller cutoff value.

\section{Summary and Conclusions}

To summarize, we have analyzed the cutoff, coupling, and momentum dependence
of the $K^{\ast }$ contributions to the nucleon's vector strangeness content
in the hadronic one--loop model of Ref. \cite{bar98}. It turns out that the
softer $K^{\ast }NY$ form factors now generally favored by $NY$ potential
models have some welcome consequences for such models. First, they can
reduce the $K^{\ast }$ contributions to the strangeness radius and magnetic
moment by over an order of magnitude, thereby indicating that the
contributions from the lightest $KY$ intermediate states might be sufficient
for rough estimates of the strangeness content in one--loop models. Although
the $K^{\ast }$ contributions remain non--negligible (in particular towards
larger momentum transfers) and worsen the agreement with the data of the
HAPPEX experiment, they cease to be unrealistically large and they do not
anymore exclude some sort of (slow) ``convergence'' of the intermediate
state sum. The very sensitive dependence of the results on the $K^{\ast }$
cutoff emphasizes, however, the limited and mostly\ qualitative character of
hadron--loop model predicitions.

H.F. would like to thank the National Institute for Nuclear Theory in
Seattle for hospitality during the 1998 Strangeness Program, where the plan
for this work originated. He would also like to thank Wally Melnitchouk for
useful correspondence on the Bonn--J\"{u}lich potential model, E.
Kolomeitsev for pointing out Ref. \cite{fri99}, and the Deutsche
Forschungsgemeinschaft for support under habilitation grant Fo 156/2--1.
F.S.N. and M.N. would like to thank FAPESP and CNPq, Brazil, for support.

\appendix

\section{Vertex Functions}

In the loop model of section \ref{mod} the strangeness form factors receive
four distinct types of contributions. Three of them correspond to amplitudes
associated with processes in which the current couples either to the hyperon
line (Y), the meson line (M), or the meson-baryon vertex (V) in the loop.
The fourth contribution corresponds to the amplitude associated with the
strangeness-current induced spin-flip transition form $K$ to $K^{\ast }$.
The corresponding vertex functions are

\begin{eqnarray}
\Gamma^{(Y)}_\mu(p^\prime,p)& =& i Q_Y \int \frac{d^4k} {(2\pi)^4} \left[%
g^2_v(F(k^2))^2 D^{\alpha\beta}(k)\left(\gamma_\alpha +i{\frac{\kappa}{2m_N}}
\sigma_{\alpha\nu}k^\nu\right) S(p^\prime-k) \gamma_\mu S(p-k) \times \right.
\nonumber \\
*[7.2pt] &&\left.\left(\gamma_\beta-i{\frac{\kappa}{2m_N}}%
\sigma_{\beta\gamma} k^\gamma\right)-g_{ps}^2(F_K(k^2))^2\Delta(k^2)\gamma_5
S(p^\prime-k) \gamma_\mu S(p-k)\gamma_5\right]\; ,  \label{1B}
\end{eqnarray}
\begin{eqnarray}
\Gamma^{(M)}_\mu(p^\prime,p)& =&- i Q_{M} \int \frac{d^4k}{(2\pi)^4} \left[
g^2_vF((k+q)^2)F(k^2) D^{\alpha\lambda}(k+q)
D^{\sigma\beta}(k)\left(\gamma_\alpha + \right.\right.  \nonumber \\
*[7.2pt] &+&\left.i{\frac{\kappa}{2m_N}} \sigma_{\alpha\nu}(k+q)^\nu\right)
[(2k+q)_\mu \, g_{\sigma\lambda}-(k+q)_\sigma g_{\lambda\mu}-k_\lambda
g_{\sigma\mu}]\times  \nonumber \\
*[7.2pt] &&S(p-k)\left(\gamma_\beta-i{\frac{\kappa}{2m_N}}%
\sigma_{\beta\gamma} k^\gamma\right)+g_{ps}^2F_K((k+q)^2)F_K(k^2) \times 
\nonumber \\
*[7.2pt] &&\left.\Delta((k+q)^2)\Delta(k^2)(2k+q)_\mu\gamma_5S(p-k)\gamma_5 
\right]\; ,  \label{1M}
\end{eqnarray}
\begin{eqnarray}
\Gamma^{(V)}_\mu(p^\prime,p)& =& Q_{M} \int \frac{d^4k} {(2\pi)^4}
\left\{g^2_v F(k^2) D^{\alpha\beta}(k) \left[i \left(\frac{ (q+2k)_\mu}{
(q+k)^2-k^2} \left(F(k^2)\, - F((k+q)^2)\right) \times \right.\right.\right.
\nonumber \\
*[7.2pt] & & \left(\gamma_\alpha +i{\frac{\kappa}{2m_N}}\sigma_{\alpha\nu}k^%
\nu\right) S(p-k)\left(\gamma_\beta-i{\frac{\kappa}{2m_N}}%
\sigma_{\beta\gamma}k^\gamma\right) - \frac{ (q-2k)_\mu}{ (q-k)^2-k^2}
(F(k^2)+  \nonumber \\
*[7.2pt] &-&\left.F((k-q)^2))\left(\gamma_\alpha +i{\frac{\kappa}{2m_N}}%
\sigma_{\alpha\nu} k^\nu\right) S(p^\prime-k)\left(\gamma_\beta-i{\frac{%
\kappa}{2m_N}}\sigma_{\beta\gamma}k^\gamma \right) \right) \; +  \nonumber \\
*[7.2pt] &+&\;{\frac{\kappa}{2m_N}}\left(F((k+q)^2)\sigma_{\alpha\mu}
S(p-k)\left(\gamma_\beta-i{\frac{\kappa}{2m_N}}\sigma_{\beta\gamma}k^\gamma
\right)\right. +  \nonumber \\
*[7.2pt] &-&\left.\left.F((k-q)^2)\left(\gamma_\alpha +i{\frac{\kappa}{2m_N}}
\sigma_{\alpha\nu}k^\nu \right)S(p^\prime-k)\sigma_{\beta\mu}\right)\right]%
-ig_{ps}^2 F_K(k^2) \Delta(k^2)\times  \nonumber \\
*[7.2pt] &&\left[\frac{ (q+2k)_\mu}{ (q+k)^2-k^2} \left(F_K(k^2)\, -
F_K((k+q)^2)\right)\gamma_5 S(p-k)\gamma_5\right.  \nonumber \\
*[7.2pt] &-& \left.\left.\frac{ (q-2k)_\mu}{ (q-k)^2-k^2} \left(F_K(k^2)-
F_K((k-q)^2)\right) \gamma_5S(p^\prime-k)\gamma_5\right]\right\} \; ,
\label{1V}
\end{eqnarray}
\begin{eqnarray}
\Gamma^{(K/K^*)}_\mu(p^\prime,p)& =&-{\frac{g_vg_{ps}F_{K^*K}^{(s)}(0)}{%
m_{K^*}}} \epsilon_{\mu\nu\lambda\alpha}\int \frac{d^4k}{(2\pi)^4}%
\left\{F((k+q)^2)F_K (k^2)D^{\alpha\beta}(k+q)\times\right.  \nonumber \\
*[7.2pt] &&\Delta(k^2)(k+q)^\nu k^\lambda \left(\gamma_\beta +i{\frac{\kappa%
}{2m_N}} \sigma_{\beta\delta}(k+q)^\delta\right)S(p-k)\gamma_5+  \nonumber \\
*[7.2pt] &+&F(k^2)F_K((k+q)^2)D^{\alpha\beta}(k)\Delta((k+q)^2)k^\nu
(k+q)^\lambda \gamma_5 \times  \nonumber \\
*[7.2pt] &&\left.S(p-k)\left(\gamma_\beta -i{\frac{\kappa}{2m_N}}%
\sigma_{\beta\delta} k^\delta\right)\right\}\; .  \label{1KK}
\end{eqnarray}

\noindent In the above equations we define $p^\prime=p+q$ and use the
notation $D_{\alpha\beta}(k)=(-g_{\alpha\beta} + k_\alpha
k_\beta/m_{K^*}^2)(k^2-m_{K^*}^2+i\epsilon)^{-1}$ for the $K^*$ propagator, $%
\Delta(k^2)=(k^2-m_K^2+i\epsilon)^{-1}$ for the kaon propagator, $S(p-k) = (p%
\kern-.5em\rlap/{\,}- \, k\kern-.5em\rlap/{\,}-\,m_Y+ i\epsilon)^{-1}$ for
the hyperon, $Y$, propagator with mass $m_\Lambda=1116$ MeV. The strangeness
charges are $Q_Y =1$ and $Q_M=-1$ and $F(k^2)$, $F_K(k^2)$ refer
respectively to the $K^*$ and kaon form factors, both given by Eq.(\ref{ff}).

\begin{table}[bh]
\begin{tabular}{|c|c|c|c|c|}
\hline
& \multicolumn{2}{c|}{$\langle r_s^2 \rangle_D \; (\mbox{fm}^2) $} & 
\multicolumn{2}{c|}{$\mu_s \; $} \\ \hline
$\Lambda$ (GeV) & 1.2 & 2.2 & 1.2 & 2.2 \\ \hline
&  &  &  &  \\ 
$KK \Lambda$ & $-0.007$ &  & $-0.237 $ &  \\ 
$K^*K^* \Lambda$ & $0.0023$ & $0.030$ & $-0.180$ & $-4.149$ \\ 
$KK^* \Lambda$ & $0.0207$ & $0.085$ & $0.253$ & $1.023$ \\ \hline
\end{tabular}
\caption{Intermediate state contributions to the strange magnetic moment $%
\protect\mu _{s}$ and the electric strangeness radius $\langle
r_{s}^{2}\rangle _{D}$ in the loop model of \protect\cite{bar98}.}
\label{kstartab4}
\end{table}

\newpage

\begin{figure}[tbp]
\caption{The dependence of a) the Dirac strangeness radius and b) the
strangeness magnetic moment on the cut--off parameter of the $K^{\ast
}N\Lambda $ form factor. The dashed (dot--dashed) line corresponds to the $%
K^{\ast }K^{\ast }\Lambda $ ($KK^{\ast }\Lambda )$ contributions, the dotted
line represents the kaon contribution, and the full line gives the total
result.}
\label{fig1}
\end{figure}


\begin{figure}[tbp]
\caption{The dependence of a) the Dirac strangeness radius and b) the
strangeness magnetic moment on the cut--off parameter of the $K^*N\Lambda $
form factor for different values of the $K^*N\Lambda $ couplings: ($g_{v}/%
\bar{4\protect\pi },\protect\kappa $) = \{($-1.588,3.26$) solid line, ($%
-0.8,2.0$) dashed line, ($-0.8,4.0$) dot-dashed line, ($-2.0,2.0$)
long-dashed line, ($-2.0,4.0$) dotted line\}.}
\label{fig2}
\end{figure}

\begin{figure}[tbp]
\caption{ Strange magnetic form factor of the nucleon. The solid and dashed
lines give the results using $\Lambda=1.2$ GeV and $\Lambda=0.9$ GeV
respectively and the data points show the new and (for comparison) earlier 
results of the SAMPLE experiment \protect\cite{mit}.}
\label{fig3}
\end{figure}


\begin{figure}[tbp]
\caption{ Strange electric and magnetic form factor combination as measured
by the HAPPEX collaboration \protect\cite{TJN}. The solid and dashed lines give the
results using $\Lambda=1.2$ GeV and $\Lambda=0.9$ GeV respectively.}
\label{fig4}
\end{figure}

\begin{figure}[tbp]
\caption{ Strange Dirac form factor of the nucleon calculated using $%
\Lambda=1.2$ GeV. The solid, dashed and dot-dashed lines give the kaon, $K^*$
and $K/K^*$ transition contributions respectively.}
\label{fig5}
\end{figure}

\begin{figure}[tbp]
\caption{ Strange Pauli form factor of the nucleon calculated using $%
\Lambda=1.2$ GeV. The solid, dashed and dot-dashed lines give the kaon, $K^*$
and $K/K^*$ transition contributions respectively.}
\label{fig6}
\end{figure}

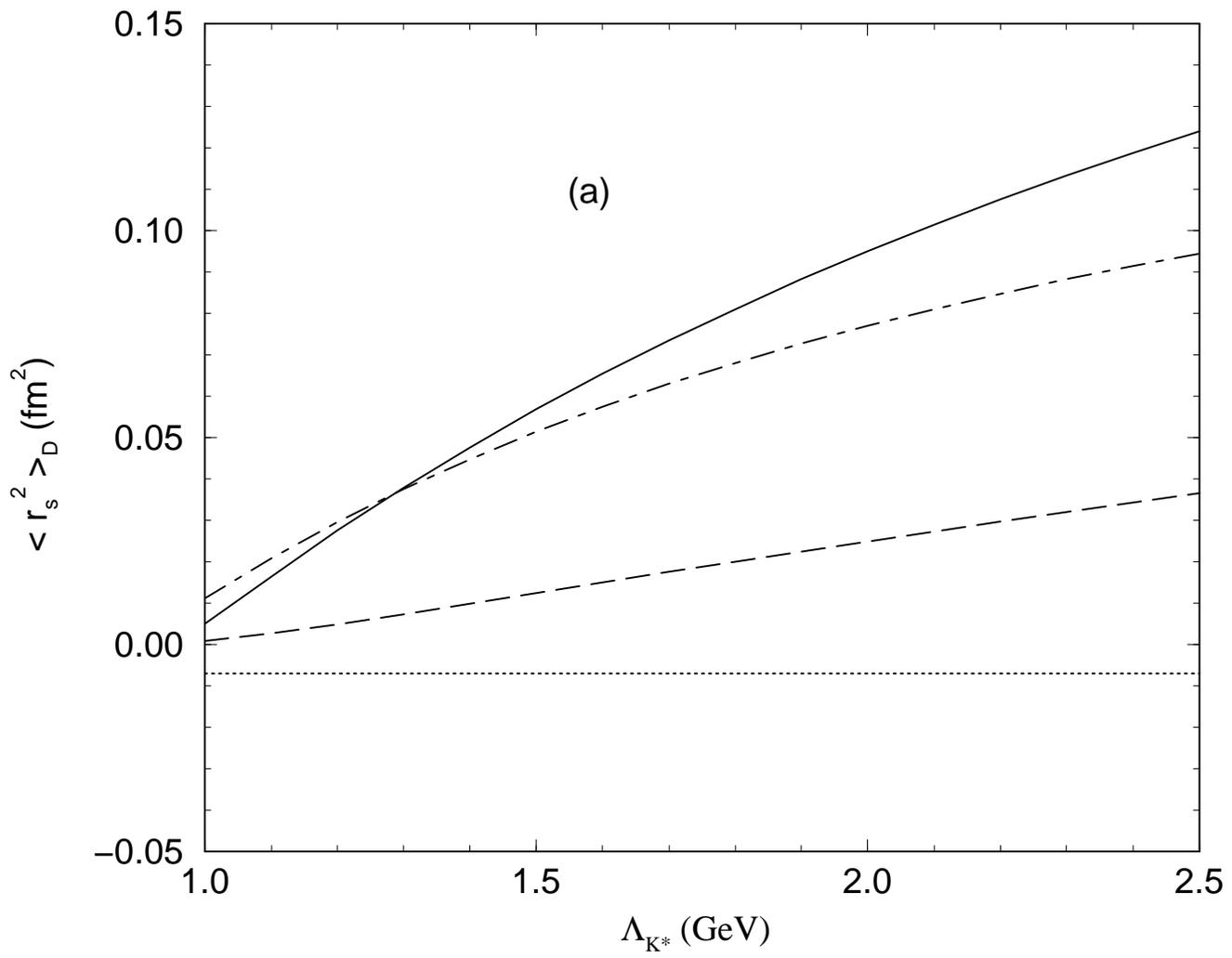
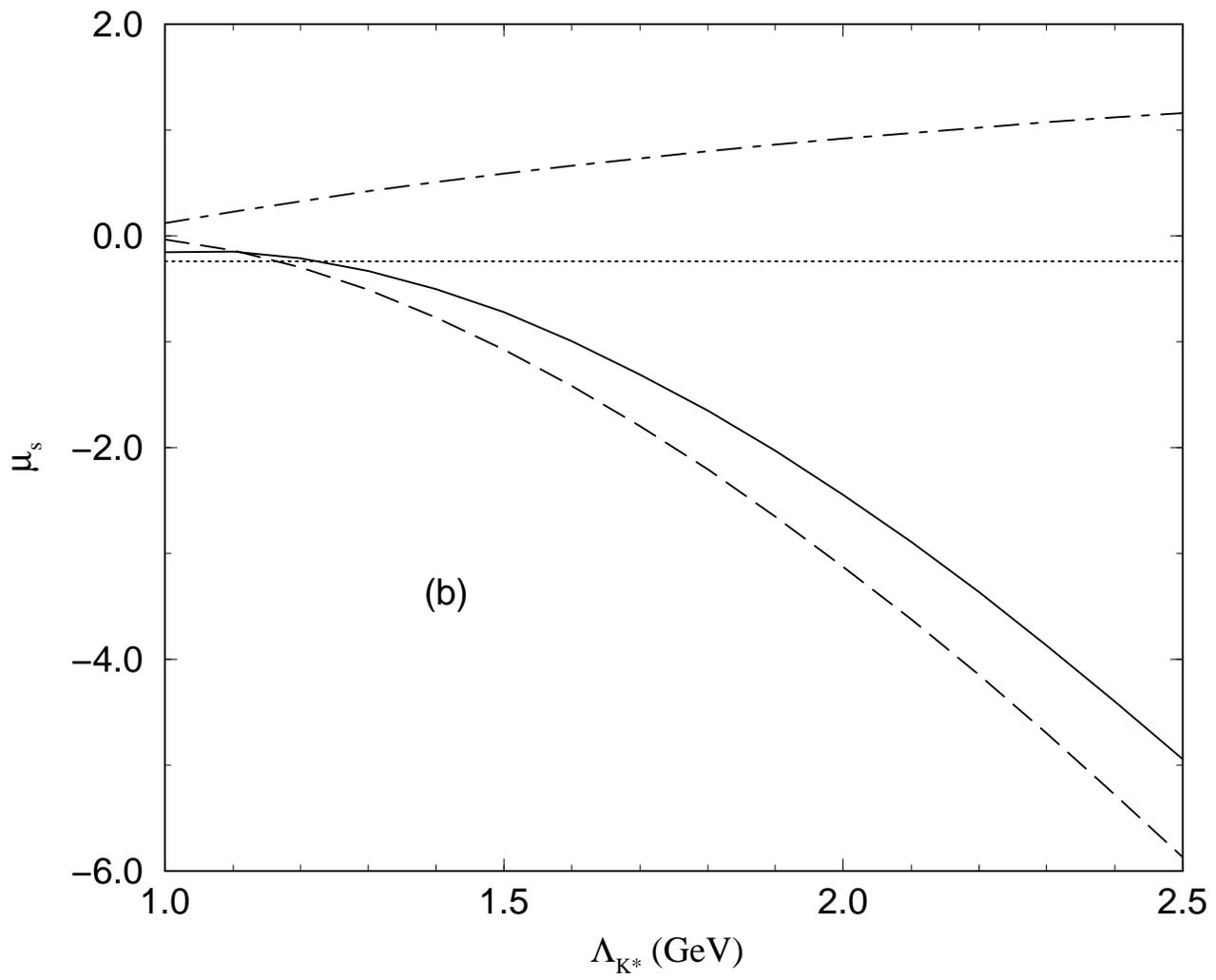
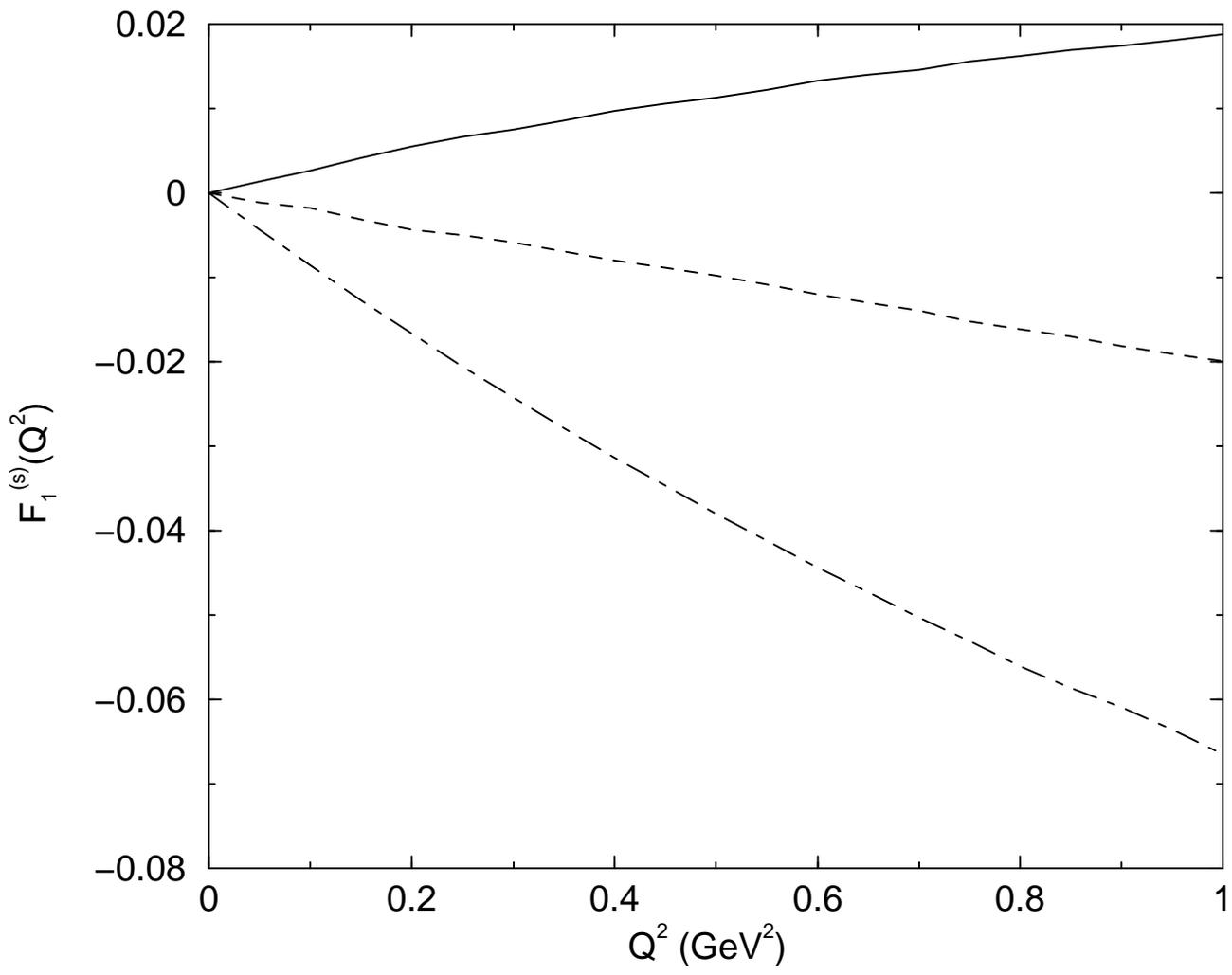
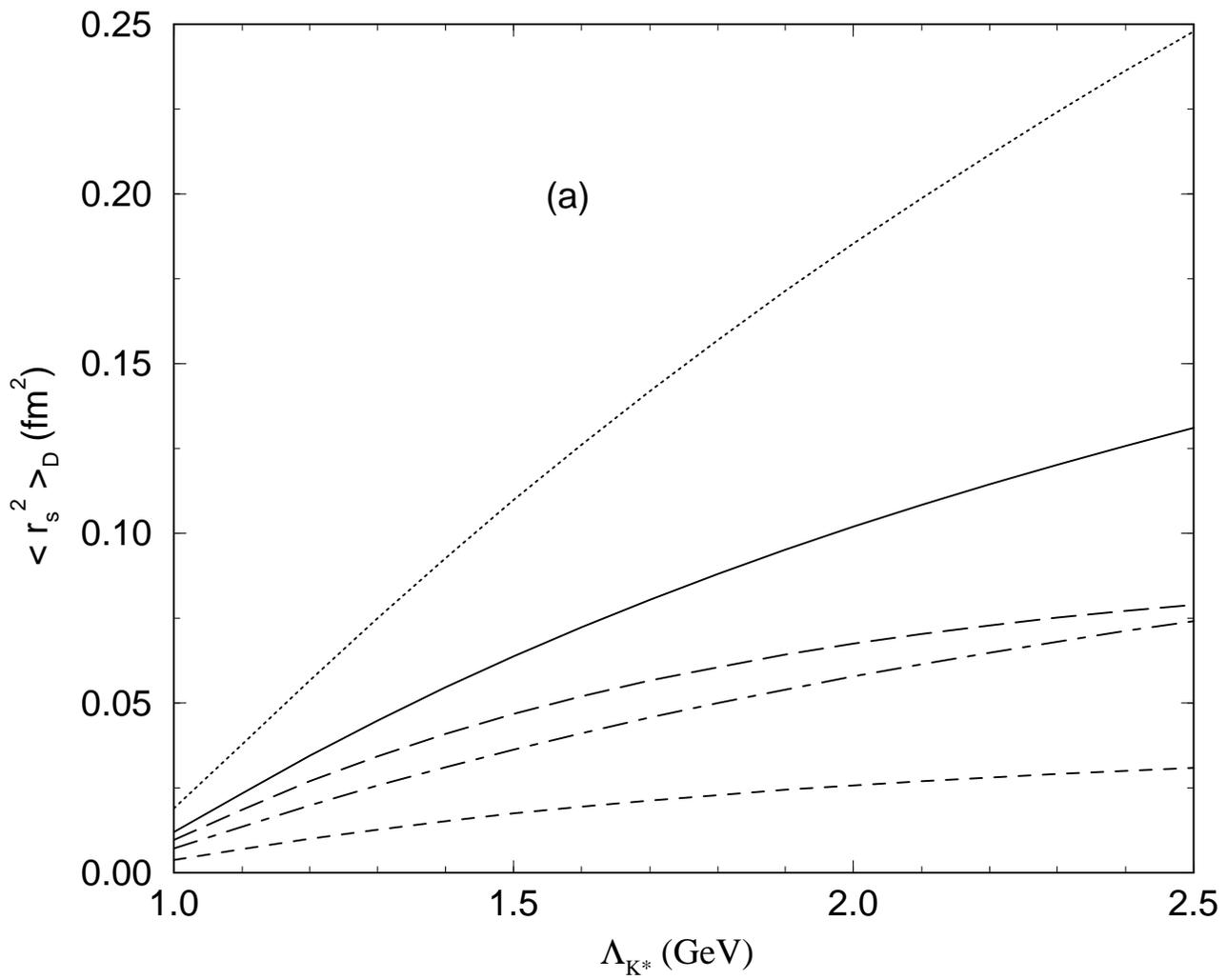
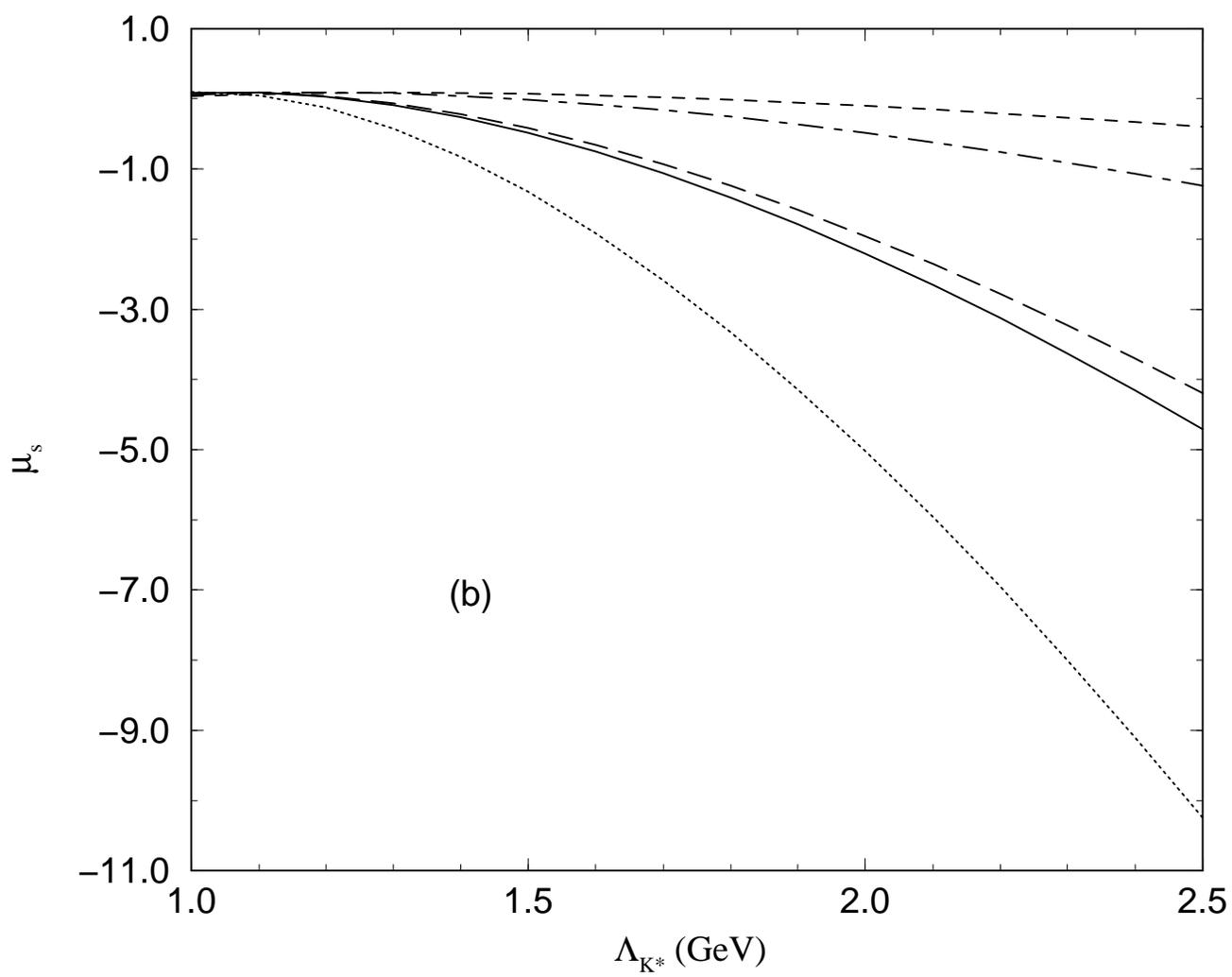
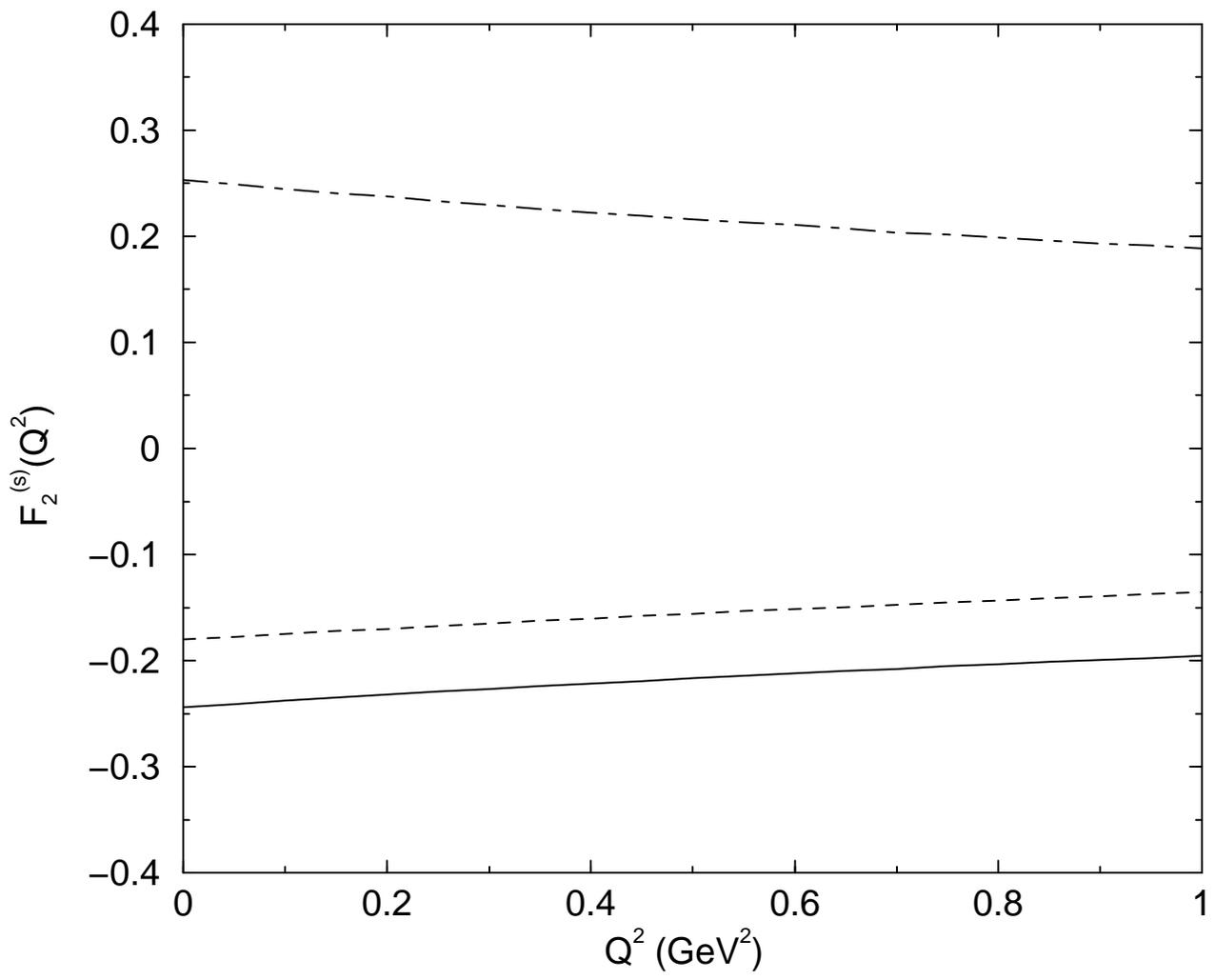
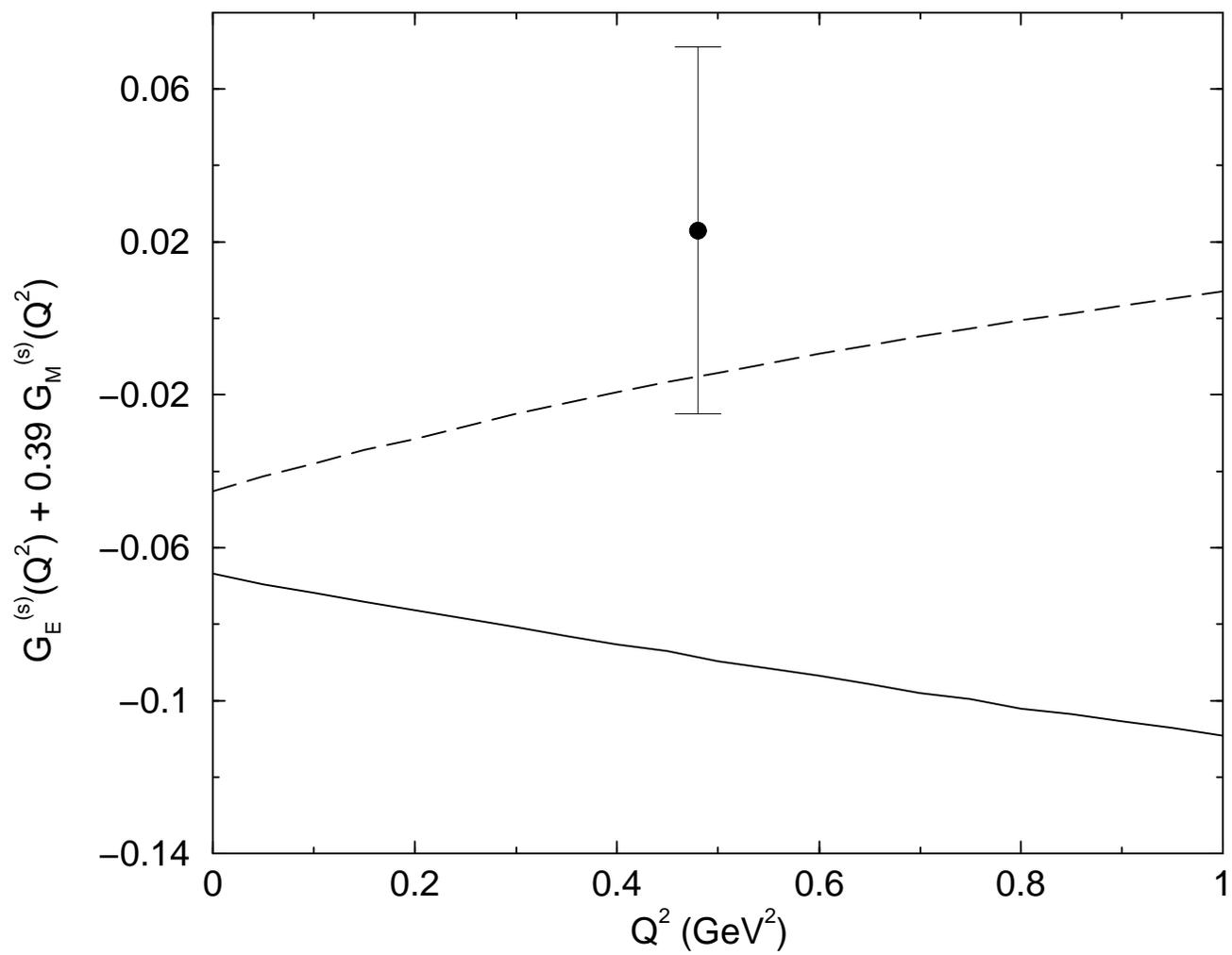
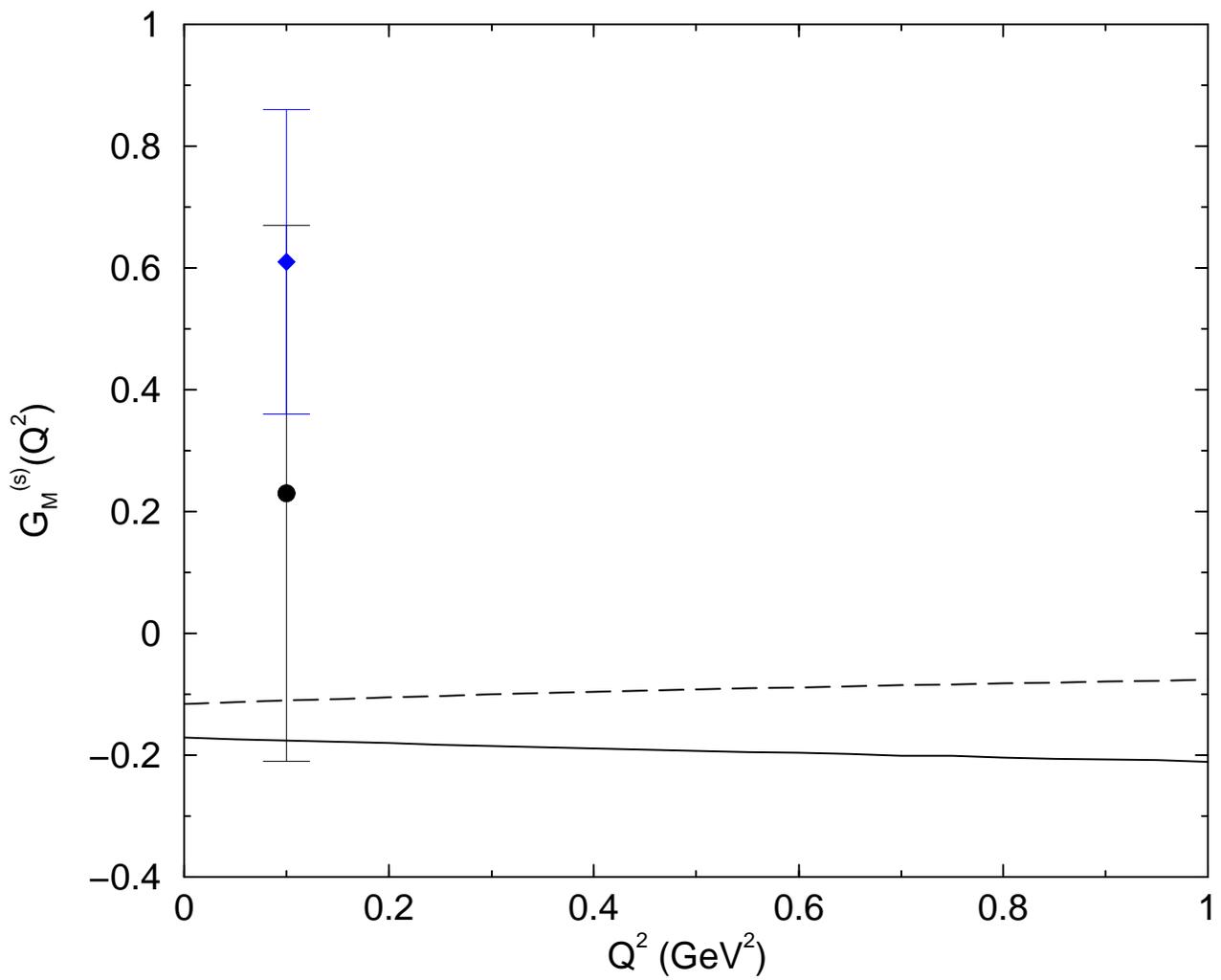
\end{document}